\documentclass[useAMS,usenatbib,usegraphicx]{mn2e}
\usepackage{epsfig}
\usepackage{amsmath} 

\usepackage{times}

\def\kms{km ${\rm s}^{-1}$}

\def\ch2{$\chi^2$}

\def\kms {\hbox{${\rm km\ s}^{-1}$}}

\def\scm  {$\hbox{{\rm cm}}^{-2}$}    

\def\MOLH {\hbox{${\rm H}_2$}}  

\def \HI {H{\sc \,i}}
\def \WpHz {W Hz$^{-1}$}

\def\lapp{\ifmmode\stackrel{<}{_{\sim}}\else$\stackrel{<}{_{\sim}}$\fi}
\def\gapp{\ifmmode\stackrel{>}{_{\sim}}\else$\stackrel{>}{_{\sim}}$\fi}

\title[\HI\ and OH in radio galaxies and quasars]{Redshifted \HI\ and OH absorption in radio galaxies and quasars}
\author[S. J. Curran et al.]{S. J. Curran$^{1}$\thanks{E-mail:
sjc@phys.unsw.edu.au}, M. T. Whiting$^{1,2}$,
M. T. Murphy$^{3,4}$, J. K. Webb$^{1}$, C. Bignell$^{5}$, 
\newauthor 
A. G. Polatidis$^{6,7}$, T. Wiklind$^{8,9,10}$, P. Francis$^{11}$
and G. Langston$^{5}$\\
$^{1}$School of Physics, University of New South Wales, Sydney NSW 2052, Australia\\
$^{2}$CSIRO Australia Telescope National Facility, PO Box 76, Epping NSW 1710, Australia\\
$^{3}$Institute of Astronomy, Madingley Road, Cambridge CB3 0HA, UK\\
$^{4}$Centre for Astrophysics and Supercomputing, Swinburne University
of Technology, PO Box 218, Hawthorn, VIC 3122, Australia\\
$^{5}$National Radio Astronomy Observatory, P.O. Box 2, Rt. 28/92
Green Bank, WV 24944-0002, USA\\
$^{6}$Max Planck Institut f\"{u}r Radioastronomie, Postfach 2024, D-53010, Bonn, Germany\\
$^{7}$Netherlands Institute for Radio Astronomy (ASTRON), Postbus 2, 7990 AA Dwingeloo, The Netherlands \\
$^{8}$Onsala Space Observatory, S-439 92 Onsala, Sweden\\
$^{9}$Space Telescope Science Institute, Baltimore, Maryland 21218, USA\\
$^{10}$Joint ALMA Observatory, Santiago, Chile\\
$^{11}$Department of Physics, Australian National University, ACT 0200, Australia
}

\begin{document}

\date{Accepted ---. Received ---; in original form ---}

\pagerange{\pageref{firstpage}--\pageref{lastpage}} \pubyear{2011}

\maketitle

\label{firstpage}

\begin{abstract}
  From a survey for redshifted \HI\ 21-cm and OH 18-cm absorption in the hosts of a sample of radio
  galaxies and quasars, we detect \HI\ in three of the ten and OH in none of the fourteen
  sources for which useful data were obtained. As expected from our recent result, all of the
  21-cm detections occur in sources with ultra-violet continuum luminosities of $L_{\rm
    UV}\leq10^{23}$ \WpHz. At these ``moderate'' luminosities, we also obtain four non-detections,
  although, as confirmed by the equipartition of detections between the type-1 and type-2 objects,
  this near-50\% detection rate cannot be attributed to unified schemes of active galactic nuclei
  (AGN). All of our detections are at redshifts of $z\lapp0.67$, which, in conjunction with our
  faint source selection, biases against UV luminous objects. The importance of ultra-violet
  luminosity (over AGN type) in the detection of 21-cm is further supported by the
  non-detections in the two high redshift ($z\sim3.6-3.8$) radio galaxies, which are both type-2
  objects, while having $L_{\rm UV} > 10^{23}$ \WpHz. Our 21-cm detections in combination with those
  previously published, give a total of eight (associated and intervening) \HI\ absorbing sources
  searched and undetected in OH. Using the detected 21-cm line strengths to normalise the limits, we find that only
  two of these eight may have been searched sufficiently deeply in OH, although even these are
  marginal.
\end{abstract}


\section{Introduction}
\label{intro}

Although opaque to optical light, the dusty Universe is transparent to
radiation at radio wavelengths, thus making the spectroscopic study of
the 21-cm spin-flip transition of neutral hydrogen (\HI) a very useful
tool in probing the far reaches of the cosmos. The low probability of
the transition compounded by the inverse square law, renders \HI\ 21-cm
currently undetectable in emission at redshifts of $z\gapp0.1$. However, in the absorption of radio
waves emitted from background quasars, the line strength depends only
upon the column density of the absorber and the flux of the background
source. Therefore by using absorption lines, we can in principal probe
\HI\ to redshifts of $z\sim50$ (or when the Universe was 1\% its
present age), where the ionosphere begins to affect low frequency
radio waves ($\lapp30$ MHz). With such observations we can address several outstanding questions in
cosmology and fundamental physics:
\begin{enumerate}

\item Probe the Epoch of Re-ionisation -- when the first ever stars
ignited, re-ionising the gas in the  smaller cosmos (e.g. \citealt{cgfo04}).

\item Determine the contribution of the neutral gas content to the mass density of the Universe (\citealt{kpec09,cur09a}).

\item Measure any putative variations in the values of the fundamental
  constants of nature at large look-back times, to at least an order
  of magnitude the sensitivity provided by the best optical data
  (see \citealt{tmw+06}). This offers one of the few experimental
  tests of current Grand Unified Theories, thus having profound
  implications for modern physics.
\end{enumerate}
This latter point requires the comparison of the redshift of the 21-cm
line with other transitions, which may be optical/ultra-violet [from
singly-ionised metals, giving $\Delta(\mu\alpha^2 g_{p})/\mu\alpha^2
g_{p}$], where $\alpha$ is the fine structure constant, $\mu$
the electron-to-proton mass ratio and $g_{p}$ the proton g-factor,
millimetre-wave [rotational transition of molecules, giving
$\Delta(\alpha^2 g_{\rm p})/\alpha^2 g_{\rm p}$] or other
decimetre transitions (see \citealt{cdk04} and references therein). Specifically, transitions
arising from the hydroxyl radical (OH), which can also be intra-compared \citep{dar03}, thus
avoiding possible line-of-sight effects which could mimic a change in the
constants.

Thus, highly redshifted \HI\ 21-cm and OH 18-cm absorbers are of great interest, although these are
currently very rare, with only 73 \HI\ 21-cm absorption systems at $z\geq0.1$ known -- 41 of which
occur in galaxies intervening the sight-lines to more distant quasars (see table 1 of
\citealt{cur09a}), with the remainder arising in the host galaxies of the quasars themselves (see
table 1 of \citealt{cw10}). In the case of OH, the situation is more dire with only five absorbers
currently known \citep{cdn99,kc02a,kcdn03,kcl+05}. Four of these were originally found
through millimetre-wave molecular absorption, although further
surveys have proven fruitless (see \citealt{cmpw03}), which we suggest is
due to the traditional optical selection of the sources: The target of choice
in many previous surveys have been damped Lyman-$\alpha$ absorption systems (DLAs), since these 
are known to contain large columns
of neutral hydrogen ($N_{\rm HI}\geq2\times10^{20}$ cm$^{-2}$, by definition) at precisely
determined redshifts. Although 19 DLAs have been detected in the \MOLH\ Lyman and Werner UV bands
(see \citealt{nlps08}, \citealt{jwpc09} and \citealt{sgp+10}), these are at molecular fractions well below the
detection thresholds of current microwave instruments \citep{cmpw03}.  Furthermore, the molecular
abundances appear to be correlated with the colour of the background quasar in that the DLAs have
molecular fractions of ${\cal F}\equiv\frac{2N_{\rm H_2}}{2N_{\rm H_2}+N_{\rm HI}}\sim10^{-7} - 0.3$
and $V-K\lapp4$, whereas the millimetre (and OH) absorbers have
molecular fractions ${\cal F}\approx0.6 - 1$ and optical--near-infrared colours of $V-K\gapp5$  (see
figure 3 of \citealt{cww09}). 
That is, not only are the radio-band absorbers redder than those of the optical-band, but there may be a
correlation between the normalised OH line strength and optical--near infrared colour
(\citealt{cwm+06}), although this requires a larger number of detections for confirmation.

These points strongly suggest that the quasar light is reddened by the
dust in the foreground absorber, which prevents the dissociation of the
molecules by the ambient UV field. From this it is apparent that in
order to detect redshifted molecular absorption with current radio
instruments, targets must be selected on the basis of their optical and
near-IR photometry, where we select the reddest objects.
However,  the obscuration responsible
for the quasar reddening could be located anywhere between us and the 
quasar redshift (the three intervening systems are the strongest
absorbers, see Sect. \ref{ohr}) and, although wide-band decimetre scans are
more efficient than at millimetre wavelengths \citep{mcw02,cwmk03},
these are very susceptible to radio frequency interference (RFI). Therefore,
in addition to our programme of using the wide-band spectrometer on the
Green Bank Telescope (GBT) to perform 200 MHz wide frequency scans
of the entire redshift space towards very red,
radio-loud objects (see \citealt{cdbw07}), we are searching for  \HI\ and OH absorption
associated with the host galaxy of the quasar. Here we add the results 
of our recently completed searches for associated absorption and 
discuss these in the context of our previous search results \citep{cwm+06,cww+08}.

\section{Observations}
\label{observations}

\subsection{Sample selection}
\label{ss}

We observed five sources with the Effelsberg telescope and eleven
with the Green Bank Telescope (with two sources, 1107--187 \& 1504--166, common to both, Table \ref{obs}),
where the \HI\ 21-cm or OH 18-cm transition fell into an available receiver band.
These targets were originally intended to form part of the
sample of \citet{cwm+06} and, as such, our targets are largely from the Parkes
Half-Jansky Flat-spectrum Sample
(PHFS, \citealt{dwf+97}). These are bright and generally compact radio
sources for which there exists comprehensive optical photometry \citep{fww00}.
From a list of sources, for which the redshifted \HI\ and OH
frequencies fell into the available bands, we targetted a specific sub-sample
for each telescope:
\begin{enumerate}
        \item For the Effelsberg telescope, our targets were selected
on the basis of their having a ``type-2'' spectrum, with narrow
emission lines, a red continuum and only weak (if any) broad emission
line components.
For these, unified schemes of AGN imply that our line-of-sight to the
nucleus is blocked by a ``dusty torus'' (which obscures the broad
line region and only allows us to view the narrow emission lines directly),
through which we may expect to detect absorption. We therefore
selected three radio galaxies (0114+074, 0454+066 \&
1555--140\footnote{One of the proposed targets, but not observed during this run.
We have, however, since detected 21-cm absorption in this with the Australia Telescope
Compact Array \citep{cwm+06}.}) which exhibit optical spectra with narrow emission lines only, 
indicating the presence of some nuclear extinction. However, 
as since shown by \citet{cww+08,cw10}, AGN type has little bearing on
whether absorption is detected  (see Sect. \ref{hir}).

    \item For the Green Bank Telescope, as per \citet{cwm+06},
    the targets were selected on the basis of their flat radio spectra
    and very red optical--near-infrared colours, properties similar to
    the (then four, now five) objects with redshifted rotational
    absorption. 
Both 0108+388 and 0500+019 have been taken from
\citet{cmr+98}, and so are known to exhibit associated H{\sc i} absorption. The
remainder were selected from the PHFS, on the basis of their optical--near-IR
photometry \citep{fww00}, in which we selected the reddest objects
in the sample, the colours of which are believed to be due to dust
\citep{wfp+95}.  
\end{enumerate}
In this paper we present the results of 14 searches for associated
\HI\ absorption and 15 searches for OH. This is in addition to the 7
\HI\ and 14 OH searched, using the Australia Telescope Compact Array
and Giant Metre-Wave Radio Telescope (GMRT), by \citet{cwm+06} [with 3
\HI\ and 5 OH searches overlapping with the present paper] and the 11
\HI\ and 7 OH high redshift ($z\geq2.9$) sources searched by
\citet{cww+08} [with no overlaps].

\subsection{Effelsberg observations}

The Effelsberg observations were performed with the 100-metre
telescope from 9--12 May 2004. We used the UHF and 21/18-cm receivers
over various bandwidths (in order to cover as wide a redshift range is
possible, while minimising RFI) over 512 channels.  System
temperatures were typically $\lapp30$ K (when RFI was
absent). Regarding each individual source:\\ 
{\bf 4C\,+06.21 (0454+066)} was observed at
a central frequency of 1185.35 MHz over a 6~MHz bandwidth, giving a
channel spacing of 3.1 \kms. Although RFI was appreciable, an
r.m.s. noise level of 67 mK was achieved over the 3.7 hour
observation, although the flux density of the source could not be
determined.\\ 
{\bf [HB89] 0114+074} was observed at a central frequency of 1058.43
MHz over a 6~MHz bandwidth, giving a channel spacing of 3.5 \kms. RFI
was severe, dominating the 3.5 hour observation.\\ 
{\bf PKS 1107--187} was observed at a central
frequency of 1112.51 MHz over a 12 MHz bandwidth, giving a spacing of
6.6 \kms. Unfortunately, RFI dominated the one hour observation. \\
Not on the original Effelsberg target list, due to severe RFI close to 887.76 MHz
when observed at Green Bank, OH was searched for in {\bf [HB89] 1504--166} over a 12 MHz
bandwidth (giving a spacing of 8.2 \kms). Again, however, RFI was
severe, allowing nothing to be salvaged from the 5.4 hour
observation.\\ 
Due to the severe RFI we were encountering at
$\lapp1.26$ GHz, with the remaining time we selected a lower redshift
target, {\bf COINS J2355+4950 (2352+495)} [in which 21-cm is detected by \citealt{vpt+03}],
where OH would be redshifted to 1345.62 MHz. We observed  over a 100 MHz
bandwidth, giving a channel spacing of 43.5 \kms, and RFI was
relatively low, allowing us to observe the source for 3.8 hours. This gave 
an r.m.s. noise level of 49 mK and a flux density equivalent to
3.504~K.\footnote{The Effelsberg telescope has a sensitivity of 1.55 K per Jy in the
UHF-band and 1.50~K per Jy in the 21/18-cm band.}\\
The data were reduced using the {\sc gildas} and {\sc xs} packages.

\subsection{Green Bank observations}
\label{gbo}

Each of the sources targetted with the Green Bank Telescope were observed for a
total of three hours with the observations being completed over
several sessions in 2004, 2008 and 2009.\footnote{Originally intended
to be completed in 2004, thus being added to the sample
of \citet{cwm+06} [Sect. \ref{ss}].} For all observations, the Prime
Focus 1 (PF1) receiver was used backed by the GBT spectrometer, with a 50
MHz band over 8\,196 lags giving a channel spacing of 6.104 kHz
(and the velocity spacings listed in Table~\ref{obs}).  Two
separate IFs were employed in order to observe both the \HI\ and OH
lines simultaneously:\\
{\bf COINS J0111+3906 (0108+388)} was observed at 851.38 (\HI) and
998.85 (OH) MHz on 7 September 2004. Both bands were RFI free over
most of the band and system temperatures were $\leq30$~K. Minimal
flagging was required, giving 2.1 hours of total integration at 851
MHz and 1.8 hours at 999 MHz.\\ 
{\bf 4C\,-00.11 (0131--001)} was observed at 765.01 MHz
(\HI) and 868.87 MHz (in order to also cover the 1612 and 1720 MHz OH
satellite lines) in three sessions over 17 July to 31 August
2008. System temperatures were $\leq40$ and $\leq30$~K, respectively,
although RFI was bad in the lower band. After flagging, 0.9 and 1.1
hours of data remained in the \HI\ and OH bands, respectively.\\
{\bf 4C\,-02.09 (0213--026)} was observed for a total of two hours at 652.16 MHz (\HI) on
31 October 2009. The system temperature was $\approx80$~K and RFI
caused around one half of the data to be completely flagged out. Of
the remaining data, only 4 MHz of the 12.5 MHz wide band used was
relatively RFI free, a range which fortunately covered the frequency
expected for the 21-cm transition (Fig. \ref{spectra} \&
Table \ref{obs}).  The OH line was searched at 765.65 MHz on 17
September 2009 for two hours over our standard 50~MHz band-width. The
system temperature was 37~K and a clean band required minimal flagging
leaving 1.7 hours of data.\\ 
{\bf PKS 0500+019} was observed at 1051.69 MHz only,
since \HI\ absorption is already detected
\citep{cmr+98}, on 2 October 2008. The system temperature was 24~K and RFI required some flagging, leaving 
one hour of data, although there remain intermittent dips across the
band.\\ 
{\bf PKS 1107--187} was observed at 949.89 and 1113.16 MHz on 31 January
2009. System temperatures were 28 and 22~K, respectively, with only
scans with wobbly band-passes and intermittent RFI spikes requiring
removal in the lower band, leaving 1.7 hours of good data. We found
an absorption feature close to 954 MHz, although this is twice as
strong in one polarisation (see Sect. \ref{det1107}) and thus
requires confirmation. The higher band was completely clean until the last the last
half of the observation, when RFI spikes started moving across the
band, leaving 1.0 hours of good data.\\ 
{\bf PKS 1430--155} was observed at
552.04 and 647.26 MHz on 15 February 2009. System temperatures were 86
and 43~K, respectively, where RFI was severe in both bands, although 8
minutes of good data could be retrieved at the higher frequency.\\
{\bf [HB89] 1504--166} was observed at 757.15 and 888.78 MHz on 30 April 2009.  The
system temperatures in the \HI\ band was 40 K, with 0.7 hours of good
data being retained.  The OH band was completely wiped out by RFI.\\
{\bf PKS 1535+004} was observed at 315.84 and 370.46 on 25 October 2004. System
temperatures were 87 and 76~K, respectively, and severe RFI meant that
both bands had to be flagged extensively, leaving 1.2 hours of
still relatively poor data.\\ 
{\bf 4C\,-01.39 (1654--020)} was observed at 475.01 and 557.23
MHz on 6 and 10 October 2008. The lower band had a system temperature
of 40~K, although it was dominated by RFI. For the upper band, the
system temperature was 115~K, but less severe RFI meant that 1.3 hours
of data could be retained.\\ 
{\bf PKS 1706+006} was first observed on 7th
September 2004 with frequencies centred on 980.27 and 1150.03 MHz
where system temperatures were 30 and 24 K, respectively, with
extensive flagging leaving 1 hour of data. Nevertheless, RFI dominated
the higher frequency and so this was re-observed on 30 May
2009. Minimal flagging was required for these observations, although
each scan exhibited a negative flux in each polarisation.\\ 
{\bf PKS 2252--089}
was searched in \HI\ on 16 July 2009 with a band centred on 844.22
MHz with a spacing of 3.052 kHz. The system temperature was 26 K, with
very little flagging of bad data required. Since the OH band (centred
on 1036.73 MHz) required the PF2 receiver, this was observed
separately for one hour on both 5 June and 29 September 2009,
at a spacing of 6.104 kHz, the system temperature being 25 K for both
observations. On each occasion, although no time dependent flagging
was required, the effect of RFI meant only the band shown in
Fig. \ref{spectra} was clean. \\
The data were reduced using the {\sc gbtidl} software.


\subsection{Other data}
\label{sect:od}

In addition to the Effelsberg and GBT observations, we include the
unpublished results of searches of other similar sources from Giant Metre-Wave Radio Telescope archival data:
\begin{itemize}
       \item ``{\em Cold gas at high redshift}'' (PI: Braun, {\tt 02RBa01, 03RBa01}),
        which yielded good data for \HI\ and OH in  4C+41.17 
        and TXS\,1243+036:\\
{\bf 4C+41.17 (B3\,0647+415)} -- for the \HI\ observations, there were 20 hours of good data from a
total of 4 observations (between August 2002 and February 2003) over
330 to 427 good baseline pairs. The OH band was searched on several
occasions, but only the observation of 3 February 2003 gave reasonable
data, of which 6.4 hours over 401 baselines were good.\\
{\bf TXS\,1243+036 (4C+03.24)} -- the \HI\ band was observed over several runs, but
only that from 30 August 2003 proves useful, with 2.3 hours of good data over
392 baseline pairs being retained. The several OH band observations also
yielded only one good run (1 September 2002), of which 5.8 hours over
405 baselines pairs were retained.
        \item ``{\em \HI\ absorption and emission in radio galaxies at
        $z\approx0.4$}'' (PI: Blake, {\tt 05CBa01}): Only one of
        the five sources searched yielded good data and a detectable
        flux, {\bf 4C+37.25 (B2\,0847+37)}. This was observed on 17 April
        2004, with 4.82 hours over 434 baseline pairs.
\end{itemize}
All of the GMRT data were reduced using the {\sc
miriad} interferometry reduction package, with a spectrum extracted from each cube.

\section{Results}
\subsection {Observational results}

\begin{figure*}
\vspace{17.0cm} 
\includegraphics{gbt_spectra/0108+388_HI.dat.ps}
\includegraphics{gbt_spectra/0108+388_OH.dat.ps}
\includegraphics{gbt_spectra/0131-001_IF0_755MHz_10kms.dat.ps}
\includegraphics{gbt_spectra/0131-001_IF1_869MHz_10kms.dat.ps}
\includegraphics{gbt_spectra/0213-026_HI_2009_10kms.dat.ps}
\includegraphics{gbt_spectra/0213-026_OH_2009_IF1_10kms.dat.ps}
\includegraphics{0454+066_1185MHz.dat.ps}
\includegraphics{gbt_spectra/0500+019_OH.dat.ps}
\includegraphics{gmrt_spectra/4c41.17_HI_merge.dat.ps}
\includegraphics{gmrt_spectra/03rba01_4c41oh_3feb03.dat.ps}
\includegraphics{gmrt_spectra/05cba01_gal_6c0850.dat.ps}
\includegraphics{gbt_spectra_prev/1107-187_IF0_949MHz_10kms.dat.ps}
\includegraphics{gbt_spectra/1107-187_OH_21dec09_10kms.dat.ps}
\includegraphics{gmrt_spectra/02rba01_1243+036_HI_final.dat.ps}
\includegraphics{gmrt_spectra/02rba01_1243_OH.dat.ps}
\includegraphics{gbt_spectra/1430-155_OH.dat.ps}
\includegraphics{gbt_spectra_4aug09/1504-166_HI.dat.ps}
\includegraphics{gbt_spectra_prev/1535+004_IF0_370MHz_10kms_bline13.dat.ps}
\includegraphics{gbt_spectra_4aug09/1654-020_OH_3.28kms.dat.ps}
\includegraphics{gbt_spectra_prev/1706+006_HI_2004.dat.ps}
\includegraphics{gbt_spectra/1706+006_OH_blined.dat.ps}
\includegraphics{gbt_spectra/2252-090_HI.dat.ps}
\includegraphics{gbt_spectra/2252-090_OH2_IF1_2009_10kms.dat.ps} 
\includegraphics{2352+495_1346MHz.dat.ps}
\caption{The useable spectra before baseline removal.
The GBT (in blue) and GMRT (in green) spectra have the ordinate in flux density [Jy] and
are shown over the RFI-free ranges (as quoted in Table \ref{res}) and the
Effelsberg spectra (in red) have the ordinate in antenna temperature
$T_{\rm A}^*$ [K] and are shown over the whole observed band. The
abscissa the barycentric frequency [MHz]. Each is shown at a
resolution of 10 \kms\ (except for 1654--020 shown at the observed 3.3 \kms\ and
2352+495 at 44 \kms). The 1535+004 and 1706+006 OH spectra after baseline removal are shown (Sect. \ref{gbo}).
Note that the flux density for 4C\,+37.25, which is resolved, is over the central beam only.
}

\label{spectra}
\end{figure*}
\begin{figure*}
\vspace{29.0 cm} \setlength{\unitlength}{1in} 
\begin{picture}(0,0)
\put(-4,2){\includegraphics{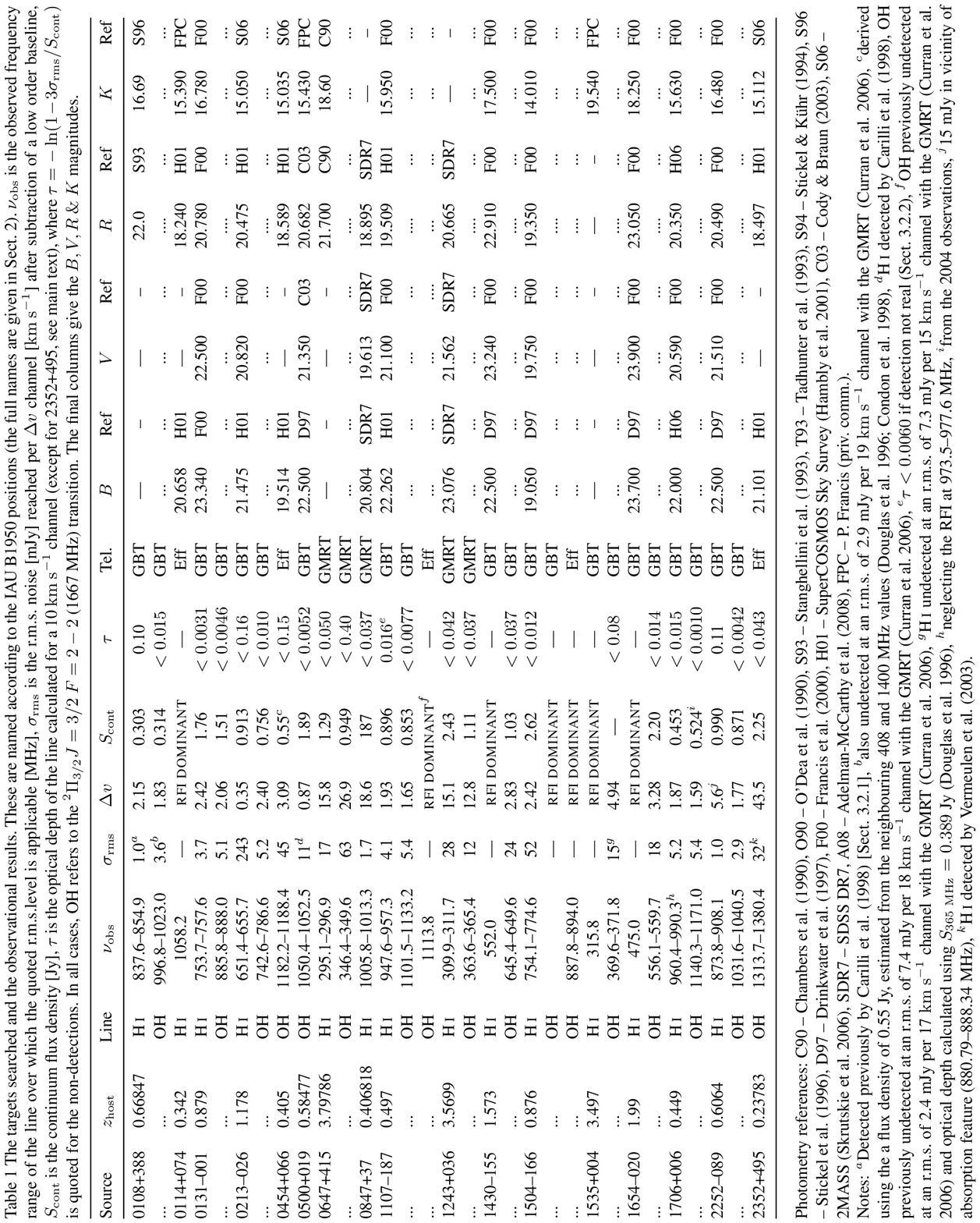}}
\end{picture}
\label{obs}
\end{figure*}
In Fig. \ref{spectra} we show the final spectra and summarise these in Table~\ref{obs}, where 
the optical depth limits are quoted per 10 \kms\ channel, apart from
the low resolution observation of 2352+495, which is quoted per
observed channel.
We have detected \HI\ in two (possibly three) 
of the targets (one of which is a re-detection) and do not detect OH
in any.

\subsection{\HI\ detections}
\label{hidet}

\subsubsection{0108+388}
\label{det0108}
\begin{figure}
\vspace{6.0 cm} \setlength{\unitlength}{1in} 
\begin{picture}(0,0)
\put(-0.2,2.8){\includegraphics{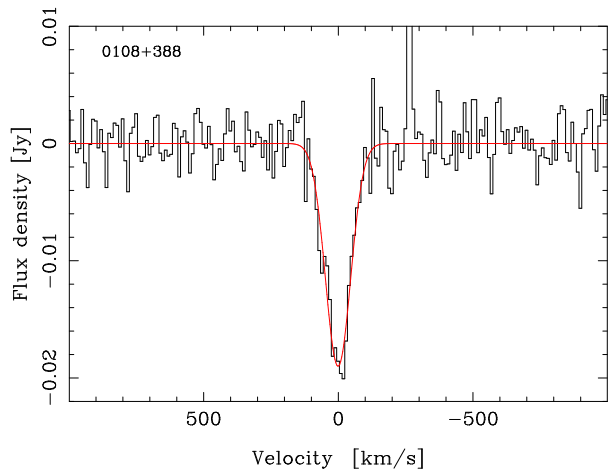}}
\end{picture}
\caption{Single Gaussian fit to the \HI\ absorption in 0108+388 shown at a resolution of 10 \kms.
The velocity scale is relative to 851.32 MHz}
\label{0108+388_gauss}
\end{figure}
In Fig. \ref{0108+388_gauss} we show the spectrum of the \HI\ 21-cm
absorption profile in 0108+388, which was previously detected by
\citet{cmr+98}. The fitting of a single Gaussian to the profile gives
a peak depth of $18.8\pm0.7$ mJy ($\approx73$ mJy by \citealt{cmr+98})
and a FWHM of $112\pm4$ \kms\ (cf. $94\pm10$ \kms) at an observed
frequency of $851.326\pm0.004$ MHz (giving a redshift
$z=0.66846-0.66847$ for the peak of the line). The previous
observations were performed with the WSRT, giving a flux density of
180 mJy (cf. our 302 mJy). We use this previously imaged, more
resolved, emission to derive an optical depth of $\tau=0.10$, which,
integrated over the FWHM of the profile, gives a column density of
$N_{\rm HI}=2.3\pm0.2\times10^{19}.\,(T_{\rm s}/f)$ \scm, where
$T_{\rm s}$ is the spin temperature of the 21-cm transition and $f$ is
the covering factor. This column density is only 30\% of the value
obtained by \citet{cmr+98} [$8.1\pm0.2\times10^{19}.\,(T_{\rm s}/f)$
\scm], which we believe
is due to their lower quality spectrum.\footnote{$N_{\rm HI}=8.1\pm0.2\times10^{19}.\,(T_{\rm s}/f)$
  \scm\ is the value derived using a Gaussian fit, whereas summing  the actual channels over which the absorption occurs gives
 $8.0\pm2.2 \times10^{19}.\,(T_{\rm s}/f)$ \scm. Applying this summing
 to our profile gives $2.2\pm0.2 \times10^{19}.\,(T_{\rm s}/f)$ \scm,
 i.e. the same as the Gaussian fit.}

\subsubsection{1107--187}
\label{det1107}
\begin{figure}
\vspace{6.0 cm} \setlength{\unitlength}{1in} 
\begin{picture}(0,0)
\put(-0.2,2.8){\includegraphics{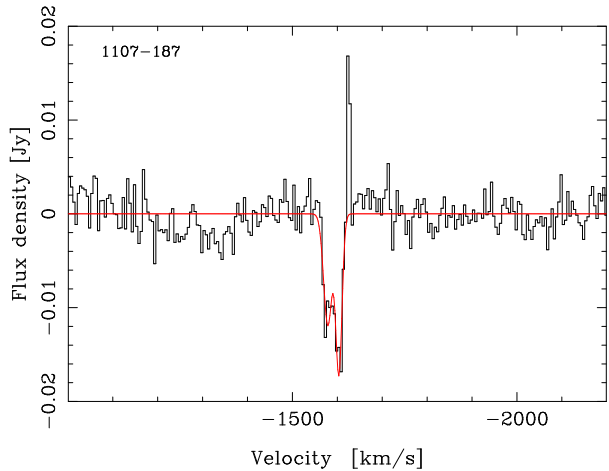}}
\end{picture}
\caption{Two Gaussian fit to the possible \HI\ absorption towards 1107--187 shown at a resolution of 5 \kms.
The velocity scale is relative to 948.83 MHz and the fits give a 17 mJy deep feature at $z=0.48905$
with a FWHM$\,=16$ \kms\ and a 12 mJy deep feature at $z=0.48917$
with a FWHM$\,=22$ \kms.}
\label{1107-187_gauss}
\end{figure}

We also report a possible 21-cm detection in 1107--187. This is apparent in
both polarisations, although with quite different depths --
$19.7\pm4.5$~mJy (at $953.86\pm0.01$ MHz with FWHM$\,=34\pm9$ \kms) in the
{\sf XX} polarisation and $9.3\pm3.1$~mJy (at $953.87\pm0.02$ MHz with
FWHM$\,=44\pm18$ \kms) in the {\sf YY} polarisation.  The average of the two
polarisations, with a single Gaussian fit, gives a line depth of
$14.4\pm1.8$~mJy with a profile width of
$50\pm11$ \kms\ (Fig.~\ref{1107-187_gauss}), the resulting optical depth of $\tau = 0.016\pm0.002$, giving
$N_{\rm HI}=1.5\pm0.5\times10^{18}.\,(T_{\rm s}/f)$ \scm. The fit is centred on
$953.87\pm0.01$ MHz, giving a redshift of $z=0.48909\pm0.00002$, cf. 0.497 quoted in the PHFS
\citep{dwf+97}. This suggests that the absorption may be blue-shifted
by $\approx 1600$ \kms\ with respect to the galaxy, although the host redshift is known to only three significant
figures. 
Note that of the many features apparent in the OH spectrum (Fig. \ref{spectra}), none give
a $^{2}\Pi_{3/2} J = 3/2$ $F=1-1$ (1665 MHz) --- $F=2-2$ (1667 MHz) pair with the same
redshift as the 21-cm line.


\subsubsection{2252--090}

\begin{figure}
\vspace{6.0 cm} \setlength{\unitlength}{1in} 
\begin{picture}(0,0)
\put(-0.2,2.8){\includegraphics{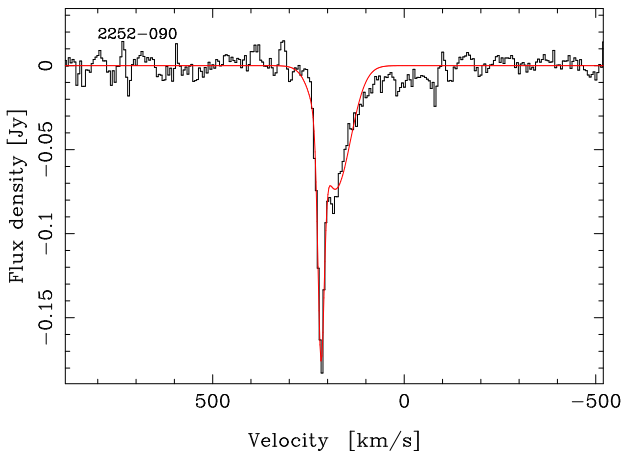}}
\end{picture}
\caption{Two Gaussian fit to the \HI\ absorption in 2252--090 shown at a resolution of 5 \kms.
The velocity scale is relative to 884.22 MHz}
\label{2252-090_gauss}
\end{figure}

Finally, we report a new detection of 21-cm absorption in 2252--090, which
was so strong as to be apparent in each 5 minute scan. This appears to
be comprised of two major components (Fig. \ref{2252-090_gauss}), one  which is narrow and deep
($\tau = 0.129\pm0.007$, FWHM$\,=17\pm1$ \kms, with $\nu = 883.578\pm0.001$
MHz giving $z = 0.60756$), as well as  a shallow, wide blue-shifted
tail ($\tau = 0.076\pm0.003$, FWHM$\,=94\pm5$ \kms, with $\nu = 883.664\pm0.008$
MHz giving $z = 0.60741\pm0.00002$).
A single Gaussian fit to the absorption gives $\tau = 0.12\pm0.01$ and
FWHM$\,=85\pm10$ \kms\ (centred on $\nu = 883.62\pm0.01$ MHz $\Rightarrow z  = 0.60748\pm0.00002$), with the
velocity integrated optical depth giving $N_{\rm
HI}=2.0\pm0.4\times10^{19}.\,(T_{\rm s}/f)$ \scm. This, along with 0108+338, is close to the maximum
line strength detected in redshifted 21-cm (see Sect. \ref{hir}, Fig. \ref{2-N-L}).

\section{Discussion}
\subsection{\HI\ results}
\label{hir}

\begin{figure*}
\centering \includegraphics[angle=270,scale=0.70]{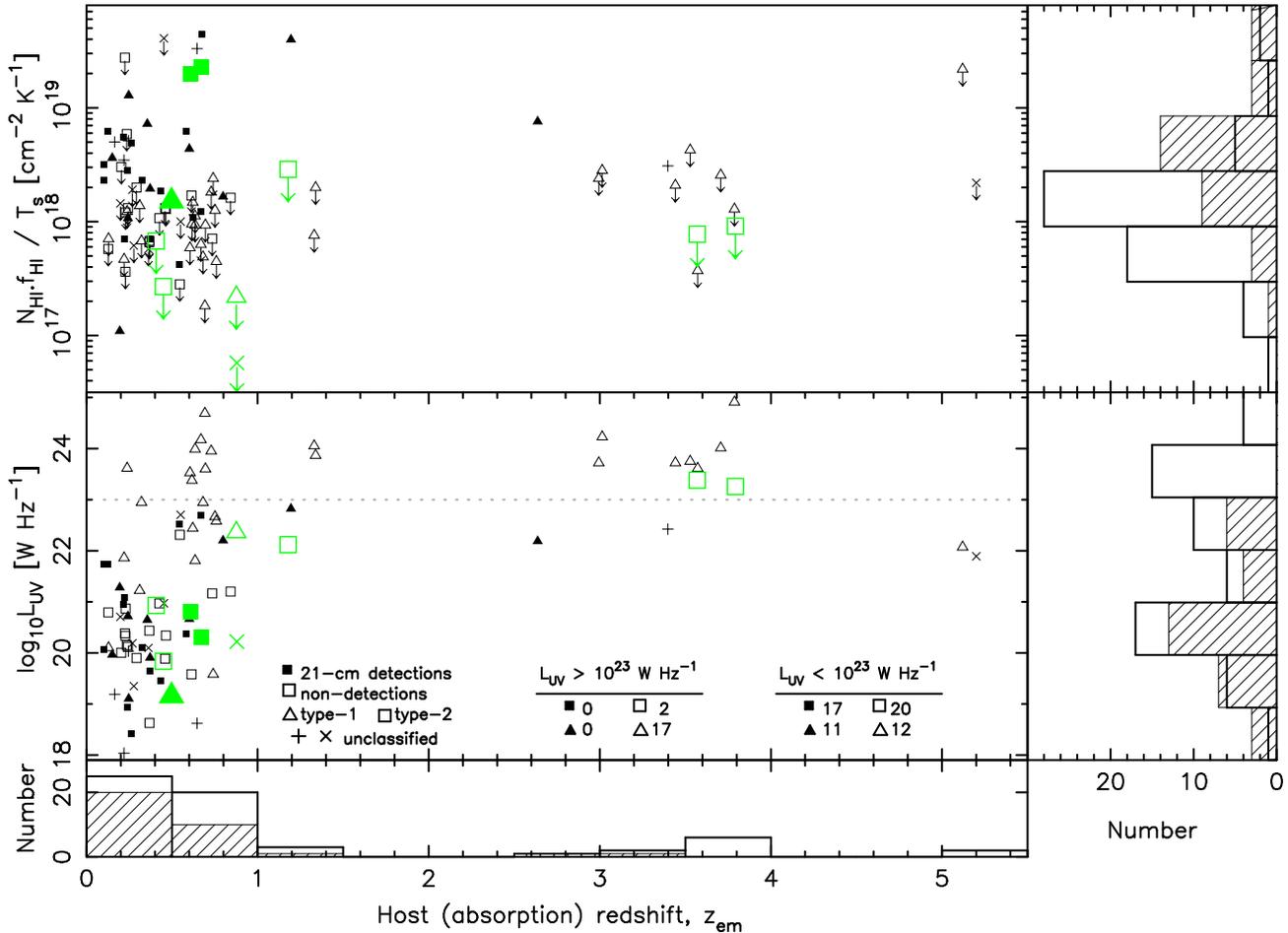}
\caption{The scaled velocity integrated optical depth of the \HI\ line
  ($1.823\times10^{18}.\int \tau dv$) [top] and the ultra-violet
($\lambda\approx1216$ \AA) luminosity [bottom] versus the host redshift for the
$z\geq0.1$ radio galaxies and quasars searched in associated 21-cm
absorption.  The filled symbols/hatched histogram represent the 21-cm
detections and the unfilled symbols/unfilled histogram the
non-detections, with the large coloured symbols designating the new
results presented here (1107--187 is treated as a detection, Sect. \ref{det1107}).
The shapes represent the AGN classifications, with
triangles representing type-1 objects and squares type-2s ({\bf +} and
{\sf x} designate an undetermined AGN type for a detection and
non-detection, respectively). The legend shows the number of each AGN
type according to the $L_{\rm UV}=10^{23}$ \WpHz\ partition.}
\label{2-N-L}
\end{figure*}

In Table \ref{res} we summarise the line strengths/limits from these new searches for redshifted \HI\ 21-cm
and OH 18-cm absorption.
\begin{table*}
\centering
\begin{minipage}{143mm}
\addtocounter{table}{1} 
\caption{The \HI\ column
densities, $N$, derived from the optical depths given in Table \ref{obs}. $T_{\rm s}$
  is the spin temperature of the \HI~21-cm, $T_{\rm x}$ is the
  excitation temperature of the OH and $f$ the
  respective covering factor. $z$-range is the redshift range over
  which the column density limit applies (Table \ref{obs}), followed
  by the galaxy/quasar classification. The final columns give the AGN
  type and 1216\AA\ luminosity [\WpHz], determined/calculated as per
  \citet{cww+08}.  }
\begin{tabular}{@{}l l l c c c c c c  @{}} 
\hline
Source  &  $z_{\rm host}$ & Line & $N$ [\scm] & $z$-range & Class & Type & Ref & $\log L_{\rm UV}$ \\
\hline
0108+388 & 0.66847  & \HI\ & $2.3\pm0.2\times10^{19}.\,(T_{\rm s}/f)$& 0.66847 & Gal & 2 & L96 & 20.309\\
...      & ...      &OH  & $<3.6\times10^{13}.\,(T_{\rm x}/f)$ & 0.62987--0.67268 &  ... &...& ... &...    \\
0131--001 &0.879   &\HI\ & $<5.7\times10^{16}.\,(T_{\rm s}/f)$ & 0.87488--0.88458 & QSO  & --& -- & 20.221\\
...      & ...      &OH  & $<1.1\times10^{13}.\,(T_{\rm x}/f)$ & 0.87766--0.88232 & ...  &...& ...&...  \\
0213--026 & 1.178   & \HI\  & $<2.9\times10^{18}.\,(T_{\rm s}/f)$  & 1.16624--1.18054 &   QSO   & 2  & D97a   &    22.119\\
...      & ...      & OH   & $<2.4\times10^{13}.\,(T_{\rm x}/f)$ &1.11960--1.24545  &...   &...& ...&...\\
0454+066  & 0.405   & OH   & $<3.6\times10^{14}.\,(T_{\rm x}/f)$ &0.40138--0.40873     &  QSO   & 2   & D97a    &  21.567 \\
0500+019 & 0.58477  & OH  &$<1.2\times10^{13}.\,(T_{\rm x}/f)$ &0.58419--0.58736  & Gal & 2 & H03  & 20.367 \\
0647+415  & 3.79786 & \HI\ &$<9.1\times10^{17}.\,(T_{\rm s}/f)$ & 3.78364--3.81379&   Gal  & 2   &  D97b & 23.258 \\
...      & ...      & OH    &$<9.5\times10^{14}.\,(T_{\rm x}/f)$ & 3.76988--3.81423&  ... &...&...& ...\\
0847+37& 0.406818 &  \HI\  & $<6.8\times10^{17}.\,(T_{\rm s}/f)$ &0.40183--0.41221 & Gal &  2& SDR7& 20.930\\
1107--187 & 0.497  &  \HI\  & $1.5\pm0.5\times10^{18}.\,(T_{\rm s}/f)^*$ & 0.48909  &   Gal  & 1   & D97a  & 19.157\\ 
...      & ...      & OH    &  $<1.8\times10^{13}.\,(T_{\rm x}/f)$ & 0.47137--0.51372&... &...&... & ..\\
1243+036 & 3.5699 & \HI\ &     $<7.7\times10^{17}.\,(T_{\rm s}/f)$  &  3.55653--3.58388    &    Gal & 2   &  R97 & 23.382  \\
 ...      & ...      & OH    & $<8.8\times10^{13}.\,(T_{\rm x}/f)$ &  3.56273--3.58607 & ... &...&... & ...\\
 1430--155& 1.573 &   OH  &      $<8.8\times10^{13}.\,(T_{\rm x}/f)$ &1.56667--1.58345 & QSO& 1 & D97a& 21.790\\
1504--166& 0.876   & \HI\  &   $<2.2\times10^{17}.\,(T_{\rm s}/f)$ & 0.83385--0.88365 &   QSO  & 1   & H78& 22.361\\
1535+004 & 3.497    & OH   &   $<1.9\times10^{14}.\,(T_{\rm x}/f)$ &3.48456--3.51125   & QSO & ---&  -- & --- \\
1654--020 &   1.99 &   OH & $<3.3\times10^{13}.\,(T_{\rm x}/f)$ & 1.97924--1.99858  &  Gal  & 1   & D97a  &  22.151   \\
1706+006 & 0.449     & \HI\ &  $<2.7\times10^{17}.\,(T_{\rm s}/f)$ & 0.43436--0.47902$^{\dagger}$  & Gal   & 2   & D97a  &19.838\\
          ...& ...&    OH  &   $<2.5\times10^{13}.\,(T_{\rm x}\/f)$ & 0.42388--0.46221   &...& ...&...&...\\
2252--090 &    0.6064  & \HI\ &  $2.0\pm0.4\times10^{19}.\,(T_{\rm s}/f)$ & $0.60748$  &  Gal  & 2  & D97a   &  20.802\\
  ...& ...&    OH      & $<1.0\times10^{13}.\,(T_{\rm x}/f)$ &0.60246--0.6162 &...& ...&... & ...\\
2352+495 & 0.23783 & OH & $<1.0\times10^{13}.\,(T_{\rm x}/f)$&0.20788--0.26921&  Gal    & 2   & L96 & 19.030  \\
\hline 
\end{tabular}
{Notes: 
$^*$$N_{\rm HI}<1.1\times10^{17}.\,(T_{\rm s}/f)$ \scm\ at $z=0.48381-0.49898$ over the absorption free region if a false detection, $^{\dagger}$with RFI at $z = 0.453-0.459$.\\References: H78 --
\citet{hms78}, L96 -- \citet{lzr+96}, D97a -- \citet{dwf+97}, D97b -- \citet{dvva97}, R97 -- \citet{rvm+97}, H03 -- \citet{hsj+03},  SDR7 -- SDSS DR7.}
\label{res}
\end{minipage}
\end{table*} 
For the 21-cm searches, we have obtained one, possibly two new
detections, as well as confirming and improving upon a previous detection. From the top
panel of Fig. \ref{2-N-L}, we see that all of the sources have been searched as 
deeply as in previous surveys and from the bottom panel, we see a range of 1216 \AA\ luminosities (given
in Table~\ref{res}): \citet{cww+08} found a critical luminosity ($L_{\rm UV}\sim10^{23}$ \WpHz) at this
wavelength, above
which 21-cm has never been detected. All but two of the sample lie below this threshold, 
these being 0647+415 and 1243+036 (Sect. \ref{sect:od}), which, as our previous $z\sim3 - 4$ searches 
\citep{cww+08}, are above the critical luminosity due to their high redshifts  
causing the selection of the brightest sources, 
despite their relatively faint magnitudes (Table \ref{obs}, cf. figure 5 of \citealt{cww09}).
These two new high redshift sources differ from the all of the other
$L_{\rm UV}\gapp10^{23}$ \WpHz\ targets searched in 21-cm
in that they are type-2 objects. Their inclusion increases the
significance of the ultra-violet luminosity effect, with the binomial probability of
 0 out of 19 detections occuring by chance being just
$1.9\times10^{-6}$, if a
21-cm detection and non-detection are equally probable. Assuming Gaussian
statistics, this corresponds to a significance of $4.76\sigma$.

As mentioned in Sect. \ref{ss}, the distribution of 21-cm detections in radio galaxies and quasars
is usually attributed to unified schemes of active galactic nuclei \citep{ant93,up95}, where, due to
the edge-on torus of dense circumnuclear material, type-2 objects (usually galaxies) present a
thick column of intervening gas along our sight-line and thus absorb in 21-cm (\citealt{jm94,cb95}),
whereas type-1 objects (usually quasars) do not.  Of the new detections, one is type-2 and one is
type-1 (assuming that 1107--187 is a detection, Sect. \ref{det1107}), and contributing to a type-2 detection
rate of 46\% and a type-1 rate of 48\%, these confirm our previous finding that unified schemes of AGN cannot be
used to explain the incidence of 21-cm absorption in these objects
(cf. \citealt{mot+01,pcv03,gss+06,gs06a}). That is, these results further support our suggestion the bulk of the cool gas
is located in the main galactic disk, which is randomly oriented with respect to the
torus of obscuring material invoked by unified schemes of AGN \citep{cw10}.

\subsection{OH results}
\label{ohr}

Although OH was not detected in any of the sample, from the \HI\ detections\footnote{0108+388,
  0500+019, 1107--187, 2352+495 (this paper),
0902+343  \citep{cb03}, J1124+1919, J1347+1217 and J2316+0404 \citep{gss+06}.} 
we can obtain normalised OH line strength limits in order to compare with the five detected OH
absorbers. 
\begin{figure*}
\centering \includegraphics[angle=270,scale=0.70]{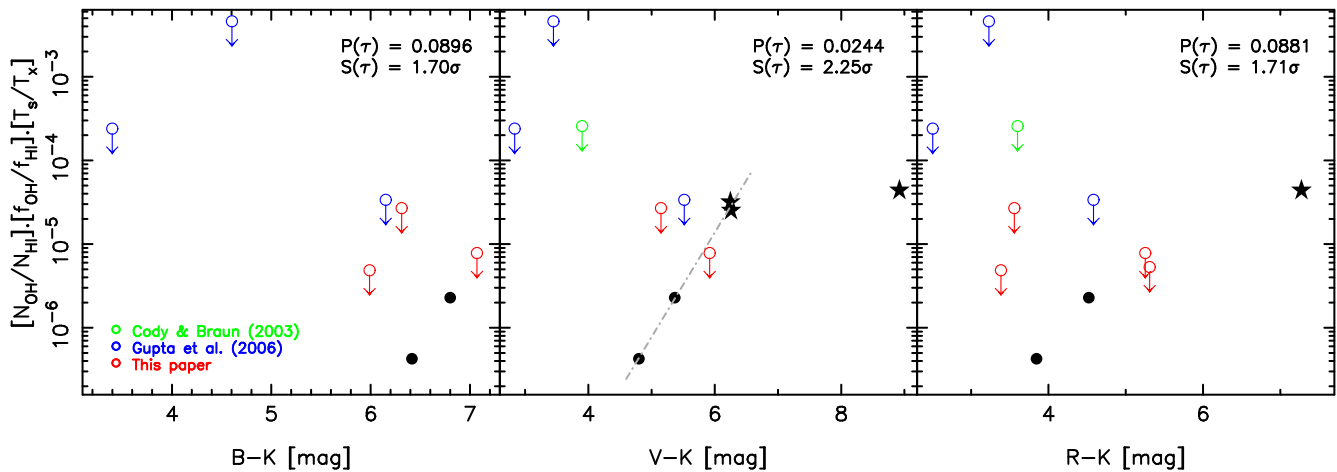}
\caption{The normalised OH $^{2}\Pi_{3/2} J = 3/2$ (1667 MHz) line
  strength ($2.38\times10^{14}\int \tau_{_{\rm OH}}\,
  dv/1.82\times10^{18}\int\tau_{_{\rm HI}}\, dv$) versus the
  blue--near-IR (left), optical--near-IR (middle) and red--near-IR
  (right) colour.  The filled symbols show the five known OH absorbers
  (circles -- associated, stars -- intervening absorbers, see
  \citealt{cwm+06} and references therein) and the unfilled symbols
  the \HI\ 21-cm detections with OH upper limits, colour coded by
  reference (green -- \citealt{cb03}, blue -- \citealt{gss+06} and red
  -- this paper).  The probability of each distribution occuring by
  chance is shown, along with the associated significance (see main
  text). The line shows the least-squares fit to those also detected
  in millimetre-band molecular transitions for the optical--near-IR
  colours \citep{cwm+06}.}
\label{OHoverH}
\end{figure*}
In Fig. \ref{OHoverH} we add the new results to the molecular line
strength/optical--near-IR colour correlation found by \citet{cwm+06},
showing also the corresponding distributions using the blue and red
magnitudes. Since the full-width half maxima (FWHM) of the OH lines are expected to be close
to those of the 21-cm profiles \citep{cdbw07}, as per \citet{cww+08}, we have rescaled the OH
column densities limits by $\sqrt{{\rm FWHM_{HI}}/\Delta v}$, in order to give the limit of 
a single channel ``smoothed'' to ${\rm FWHM_{OH} \approx FWHM_{HI}}$. This
should give a more accurate estimate of the upper limit than quoting this per $\Delta v$ channel.\footnote{$\Delta v$ is
the original resolution of the observations or the 10 \kms\ used in Table \ref{obs} and ${\rm FWHM_{HI}}$ is obtained from
\citet{mir89,cmr+98,cb03,vpt+03,gss+06}, as well as this paper (Sect. \ref{hidet}).}

From this, we see that, even if the reddening of the quasar light does occur within its host galaxy
and not at some intervening redshift, according the to the five known OH absorbers, only 0500+019 and
(from the $R-K$ plot) 0108+388 may have been searched sufficiently deeply. Although the limit is
close to the expected detection threshold, 0500+019 is also undetected in HCO$^+$, to limits
which are significantly stronger according to the $N_{\rm HCO+}$---$V-K$ correlation \citep{cwc+10},
and so perhaps the reddening of this source does not occur in the host galaxy but is the cause of
some intervening absorber.\footnote{$<30$\% of the redshift space towards 0500+019 has been scanned
  for HCO$^+$ (\citealt{mcw02}), although the detection of OH does not ensure the detection of a
  millimetre transition \citep{kcl+05}.}

Note finally, that the addition of these limits through the {\sc asurv} survival analysis package
\citep{ifn86}, increases the significance of the $V-K$ correlation over that for the OH detections
only ($S(\tau) = 1.96\sigma$, \citealt{cwm+06}). For the $B-K$ and $R-K$ correlations
the significance is somewhat lower, although, due to the limited availability of these
magnitudes for the known absorbers, this is not surprising being based upon only two 
or three detections.

\section{Summary}

We have undertaken a survey for redshifted \HI\ 21-cm and OH 18-cm absorption in a sample 
of type-2 AGN and reddened flat spectrum objects with the Effelsberg and Green Bank telescopes.
We also include unpublished searches of similar objects with the  Giant Metre-Wave Radio Telescope.
Of the ten objects for which there are useful data, we report one new clear 21-cm detection, 
in addition to  a possible detection, and confirm and significantly improve upon a previous detection (all with
the GBT). The selection criteria for the targets pre-date the findings of \citet{cww+08}, although
they confirm these:
\begin{enumerate}
   \item All of the 21-cm detections occur in objects with $\lambda = 1216$~\AA\ luminosities of $L_{\rm
       UV}\leq10^{23}$~\WpHz, a range within which there are also four new non-detections.
     All of the detections arise in galaxies (as opposed to quasars), which have been noted to have higher 21-cm detection rates 
(52\% for galaxies compared to 17\% for quasars), this being attributed to galaxies tending
     to be type-2 objects. However, \citet{cw10} have shown that the higher detection rate in
     galaxies is likely to be a consequence of their generally lower UV luminosities, as opposed
     to their AGN classification.
     
    \item The two objects for which $L_{\rm UV}>10^{23}$ \WpHz\ are undetected, confirming that the
       ultra-violet luminosity is an important criterion in the detection of cool neutral gas
       \citep{cw10}. Their inclusion, the first published type-2 $L_{\rm UV}>10^{23}$ \WpHz\ sources
         searched for in 21-cm, raises the significance of the UV luminosity---21-cm anti-correlation to $4.76\sigma$.    

  \item At $L_{\rm
       UV}\leq10^{23}$~\WpHz, the detection rates for both type-1 and type-2 objects remain close
       to 50\%, supporting the hypothesis that unified schemes of active galactic nuclei cannot
       account for the observed incidence of 21-cm absorption in radio galaxies and quasars.

\end{enumerate}
Fourteen of the sources searched in OH had useful data, although only two of these have 21-cm
detections, thus being able to yield limits on the normalised line strengths. Adding those from the
literature, gives a total of eight objects for which we can normalise the line strengths and thus
compare with the five known OH absorbers. On this basis, however, we find that only two (0108+388 \& 
0500+019, both from this paper) come close to having been searched deeply enough. This confirms
the findings of \citet{cwm+06} that many of the known objects are simply not ``red enough'' to
indicate sufficiently large columns of dust, conducive to the presence of molecular gas, along their sight-lines.

\section*{Acknowledgements}

We would like to thank Christian Henkel for the information on the Effelsberg 
telescope, Chris Blake for
his GMRT data ({\tt 05CBa01}) and Kamble Jayprakash \& Yogesh
Wadadekar for their assistance in accessing the archival GMRT data
({\tt 02RBa01 \& 03RBa01}).  Also, many thanks to Cormac ``Mopra Boy'' Purcell
for his fabulous {\sc sdfits2fits} script, which allowed easy
manipulation of the reduced GBT data, and Anant Tanna for the digitised
spectrum of 0108+388 from \citet{cmr+98}.
MTM thanks the STFC for an Advanced Fellowship and the
Australian Research Council for a QEII Research Fellowship (DP0877998).

This research has made use of the NASA/IPAC Extragalactic Database
(NED) which is operated by the Jet Propulsion Laboratory, California
Institute of Technology, under contract with the National Aeronautics
and Space Administration. This research has also made use of NASA's
Astrophysics Data System Bibliographic Services and {\sc asurv} Rev
1.2 \citep{lif92a}, which implements the methods presented
in \citet{ifn86}.

The Two Micron All Sky Survey, is a joint project of
  the University of Massachusetts and the Infrared Processing and
  Analysis Center/California Institute of Technology, funded by the
  National Aeronautics and Space Administration and the National
  Science Foundation.


\begin{thebibliography}{55}
\expandafter\ifx\csname natexlab\endcsname\relax\def\natexlab#1{#1}\fi

\bibitem[{{Adelman-McCarthy} {et~al.}(2008){Adelman-McCarthy}, {Ag{\"u}eros},
  {Allam}, {Anderson}, {Anderson}, {Annis}, {Bahcall}, \& {Baldry}}]{aaa+08}
{Adelman-McCarthy}, J.~K., {Ag{\"u}eros}, M.~A., {Allam}, S.~S., {Anderson},
  K.~S.~J., {Anderson}, S.~F., {Annis}, J., {Bahcall}, N.~A., \& {Baldry}.
  2008, ApJS, 175, 297

\bibitem[{Antonucci(1993)}]{ant93}
Antonucci, R. R.~J. 1993, ARA\&A, 31, 473

\bibitem[{{Carilli} {et~al.}(2004){Carilli}, {Gnedin}, {Furlanetto}, \&
  {Owen}}]{cgfo04}
{Carilli}, C.~L., {Gnedin}, N., {Furlanetto}, S., \& {Owen}, F. 2004, Science
  with the Square Kilometer Array, New Astronomy Reviews 48, ed. C.~L. Carilli
  \& S.~Rawlings (Amsterdam: Elsevier), 1053--1061

\bibitem[{{Carilli} {et~al.}(1998){Carilli}, {Menten}, {Reid}, {Rupen}, \&
  {Yun}}]{cmr+98}
{Carilli}, C.~L., {Menten}, K.~M., {Reid}, M.~J., {Rupen}, M.~P., \& {Yun},
  M.~S. 1998, ApJ, 494, 175

\bibitem[{{Chambers} {et~al.}(1990){Chambers}, {Miley}, \& {van
  Breugel}}]{cmv90}
{Chambers}, K.~C., {Miley}, G.~K., \& {van Breugel}, W.~J.~M. 1990, ApJ, 363,
  21

\bibitem[{{Chengalur} {et~al.}(1999){Chengalur}, {de Bruyn}, \&
  {Narasimha}}]{cdn99}
{Chengalur}, J.~N., {de Bruyn}, A.~G., \& {Narasimha}, D. 1999, A\&A, 343, L79

\bibitem[{{Cody} \& {Braun}(2003)}]{cb03}
{Cody}, A.~M. \& {Braun}, R. 2003, A\&A, 400, 871

\bibitem[{{Condon} {et~al.}(1998){Condon}, {Cotton}, {Greisen}, {Yin},
  {Perley}, {Taylor}, \& {Broderick}}]{ccg+98}
{Condon}, J.~J., {Cotton}, W.~D., {Greisen}, E.~W., {Yin}, Q.~F., {Perley},
  R.~A., {Taylor}, G.~B., \& {Broderick}, J.~J. 1998, AJ, 115, 1693

\bibitem[{{Conway} \& {Blanco}(1995)}]{cb95}
{Conway}, J.~E. \& {Blanco}, P.~R. 1995, ApJ, 449, L131

\bibitem[{Curran(2010)}]{cur09a}
Curran, S.~J. 2010, MNRAS, 402, 2657

\bibitem[{Curran {et~al.}(2007)Curran, Darling, Bolatto, Whiting, Bignell, \&
  Webb}]{cdbw07}
Curran, S.~J., Darling, J.~K., Bolatto, A.~D., Whiting, M.~T., Bignell, C., \&
  Webb, J.~K. 2007, MNRAS, 382, L11

\bibitem[{Curran {et~al.}(2004{\natexlab{a}})Curran, Kanekar, \&
  Darling}]{cdk04}
Curran, S.~J., Kanekar, N., \& Darling, J.~K. 2004{\natexlab{a}}, Science with
  the Square Kilometer Array, New Astronomy Reviews 48, ed. C.~L. Carilli \&
  S.~Rawlings (Amsterdam: Elsevier), 1095--1105

\bibitem[{Curran {et~al.}(2004{\natexlab{b}})Curran, Murphy, Pihlstr\"{o}m,
  Webb, Bolatto, \& Bower}]{cmpw03}
Curran, S.~J., Murphy, M.~T., Pihlstr\"{o}m, Y.~M., Webb, J.~K., Bolatto,
  A.~D., \& Bower, G.~C. 2004{\natexlab{b}}, MNRAS, 352, 563

\bibitem[{Curran {et~al.}(2005)Curran, Webb, Murphy, \& Kuno}]{cwmk03}
Curran, S.~J., Webb, J.~K., Murphy, M.~T., \& Kuno, N. 2005, in Highlights of
  Astronomy, Vol. 13, as presented at the XXVth General Assembly of the IAU -
  2003, ed. O.~Engvold. (San Francisco: ASP Conf. Ser.), 845 -- 847, in press
  (astro-ph/0310589)

\bibitem[{Curran {et~al.}(2006)Curran, Whiting, Murphy, Webb, Longmore,
  Pihlstr\"{o}m, Athreya, \& Blake}]{cwm+06}
Curran, S.~J., Whiting, M., Murphy, M.~T., Webb, J.~K., Longmore, S.~N.,
  Pihlstr\"{o}m, Y.~M., Athreya, R., \& Blake, C. 2006, MNRAS, 371, 431

\bibitem[{Curran \& Whiting(2010)}]{cw10}
Curran, S.~J. \& Whiting, M.~T. 2010, ApJ, 712, 303

\bibitem[{Curran {et~al.}(2010{\natexlab{a}})Curran, Whiting, Combes, Kuno,
  Francis, Nakai, Webb, Murphy, \& Wiklind}]{cwc+10}
Curran, S.~J., Whiting, M.~T., Combes, F., Kuno, N., Francis, P., Nakai, N.,
  Webb, J.~K., Murphy, M.~T., \& Wiklind, T. 2010{\natexlab{a}}, MNRAS, in
  preparation

\bibitem[{Curran {et~al.}(2010{\natexlab{b}})Curran, Whiting, \& Webb}]{cww09}
Curran, S.~J., Whiting, M.~T., \& Webb, J.~K. 2010{\natexlab{b}}, Proceedings
  of Science, accepted (arXiv:0910.3743)

\bibitem[{Curran {et~al.}(2008)Curran, Whiting, Wiklind, Webb, Murphy, \&
  Purcell}]{cww+08}
Curran, S.~J., Whiting, M.~T., Wiklind, T., Webb, J.~K., Murphy, M.~T., \&
  Purcell, C.~R. 2008, MNRAS, 391, 765

\bibitem[{Darling(2003)}]{dar03}
Darling, J. 2003, PhRvL, 91, 011301

\bibitem[{{Dey} {et~al.}(1997){Dey}, {van Breugel}, {Vacca}, \&
  {Antonucci}}]{dvva97}
{Dey}, A., {van Breugel}, W., {Vacca}, W.~D., \& {Antonucci}, R. 1997, ApJ,
  490, 698

\bibitem[{{Douglas} {et~al.}(1996){Douglas}, {Bash}, {Bozyan}, {Torrence}, \&
  {Wolfe}}]{dbb+96}
{Douglas}, J.~N., {Bash}, F.~N., {Bozyan}, F.~A., {Torrence}, G.~W., \&
  {Wolfe}, C. 1996, AJ, 111, 1945

\bibitem[{{Drinkwater} {et~al.}(1997){Drinkwater}, {Webster}, {Francis},
  {Condon}, {Ellison}, {Jauncey}, {Lovell}, {Peterson}, \& {Savage}}]{dwf+97}
{Drinkwater}, M.~J., {Webster}, R.~L., {Francis}, P.~J., {Condon}, J.~J.,
  {Ellison}, S.~L., {Jauncey}, D.~L., {Lovell}, J., {Peterson}, B.~A., \&
  {Savage}, A. 1997, MNRAS, 284, 85

\bibitem[{{Francis} {et~al.}(2000){Francis}, {Whiting}, \& {Webster}}]{fww00}
{Francis}, P.~J., {Whiting}, M.~T., \& {Webster}, R.~L. 2000, PASA, 17, 56

\bibitem[{{Gupta} \& {Saikia}(2006)}]{gs06a}
{Gupta}, N. \& {Saikia}, D.~J. 2006, MNRAS, 370, 738

\bibitem[{{Gupta} {et~al.}(2006){Gupta}, {Salter}, {Saikia}, {Ghosh}, \&
  {Jeyakumar}}]{gss+06}
{Gupta}, N., {Salter}, C.~J., {Saikia}, D.~J., {Ghosh}, T., \& {Jeyakumar}, S.
  2006, MNRAS, 373, 972

\bibitem[{{Hambly} {et~al.}(2001){Hambly}, {MacGillivray}, {Read}, {Tritton},
  {Thomson}, {Kelly}, {Morgan}, {Smith}, {Driver}, {Williamson}, {Parker},
  {Hawkins}, {Williams}, \& {Lawrence}}]{hmr+01}
{Hambly}, N., {MacGillivray}, H., {Read}, M., {Tritton}, S., {Thomson}, E.,
  {Kelly}, B., {Morgan}, D., {Smith}, R., {Driver}, S., {Williamson}, J.,
  {Parker}, Q., {Hawkins}, M., {Williams}, P., \& {Lawrence}, A. 2001, MNRAS,
  326, 1279

\bibitem[{{Hook} {et~al.}(2003){Hook}, {Shaver}, {Jackson}, {Wall}, \&
  {Kellermann}}]{hsj+03}
{Hook}, I.~M., {Shaver}, P.~A., {Jackson}, C.~A., {Wall}, J.~V., \&
  {Kellermann}, K.~I. 2003, A\&A, 399, 469

\bibitem[{{Hunstead} {et~al.}(1978){Hunstead}, {Murdoch}, \&
  {Shobbrook}}]{hms78}
{Hunstead}, R.~W., {Murdoch}, H.~S., \& {Shobbrook}, R.~R. 1978, MNRAS, 185,
  149

\bibitem[{{Isobe} {et~al.}(1986){Isobe}, {Feigelson}, \& {Nelson}}]{ifn86}
{Isobe}, T., {Feigelson}, E., \& {Nelson}, P. 1986, ApJ, 306, 490

\bibitem[{{Jaffe} \& {McNamara}(1994)}]{jm94}
{Jaffe}, W. \& {McNamara}, B.~R. 1994, ApJ, 434, 110

\bibitem[{{Jorgenson} {et~al.}(2009){Jorgenson}, {Wolfe}, {Prochaska}, \&
  {Carswell}}]{jwpc09}
{Jorgenson}, R.~A., {Wolfe}, A.~M., {Prochaska}, J.~X., \& {Carswell}, R.~F.
  2009, ApJ, 704, 247

\bibitem[{{Kanekar} {et~al.}(2005){Kanekar}, {Carilli}, {Langston}, {Rocha},
  {Combes}, {Subrahmanyan}, {Stocke}, {Menten}, {Briggs}, \&
  {Wiklind}}]{kcl+05}
{Kanekar}, N., {Carilli}, C.~L., {Langston}, G.~I., {Rocha}, G., {Combes}, F.,
  {Subrahmanyan}, R., {Stocke}, J.~T., {Menten}, K.~M., {Briggs}, F.~H., \&
  {Wiklind}, T. 2005, PhRvL, 95, 261301

\bibitem[{{Kanekar} \& {Chengalur}(2002)}]{kc02a}
{Kanekar}, N. \& {Chengalur}, J.~N. 2002, A\&A, 381, L73

\bibitem[{{Kanekar} {et~al.}(2003){Kanekar}, {Chengalur}, {de Bruyn}, \&
  {Narasimha}}]{kcdn03}
{Kanekar}, N., {Chengalur}, J.~N., {de Bruyn}, A.~G., \& {Narasimha}, D. 2003,
  MNRAS, 345, L7

\bibitem[{{Kanekar} {et~al.}(2009){Kanekar}, {Prochaska}, {Ellison}, \&
  {Chengalur}}]{kpec09}
{Kanekar}, N., {Prochaska}, J.~X., {Ellison}, S.~L., \& {Chengalur}, J.~N.
  2009, MNRAS, 396, 385

\bibitem[{{Lavalley} {et~al.}(1992){Lavalley}, {Isobe}, \&
  {Feigelson}}]{lif92a}
{Lavalley}, M.~P., {Isobe}, T., \& {Feigelson}, E.~D. 1992, in BAAS, Vol.~24,
  839--840

\bibitem[{{Lawrence} {et~al.}(1996){Lawrence}, {Zucker}, {Readhead}, {Unwin},
  {Pearson}, \& {Xu}}]{lzr+96}
{Lawrence}, C.~R., {Zucker}, J.~R., {Readhead}, A.~C.~S., {Unwin}, S.~C.,
  {Pearson}, T.~J., \& {Xu}, W. 1996, ApJS, 107, 541

\bibitem[{{Mirabel}(1989)}]{mir89}
{Mirabel}, I.~F. 1989, ApJ, 340, L13

\bibitem[{{Morganti} {et~al.}(2001){Morganti}, {Oosterloo}, {Tadhunter}, {van
  Moorsel}, {Killeen}, \& {Wills}}]{mot+01}
{Morganti}, R., {Oosterloo}, T.~A., {Tadhunter}, C.~N., {van Moorsel}, G.,
  {Killeen}, N., \& {Wills}, K.~A. 2001, MNRAS, 323, 331

\bibitem[{Murphy {et~al.}(2003)Murphy, Curran, \& Webb}]{mcw02}
Murphy, M.~T., Curran, S.~J., \& Webb, J.~K. 2003, MNRAS, 342, 830

\bibitem[{{Noterdaeme} {et~al.}(2008){Noterdaeme}, {Ledoux}, {Petitjean}, \&
  {Srianand}}]{nlps08}
{Noterdaeme}, P., {Ledoux}, C., {Petitjean}, P., \& {Srianand}, R. 2008, A\&A,
  481, 327

\bibitem[{{O'Dea} {et~al.}(1990){O'Dea}, {Baum}, \& {Morris}}]{obm90}
{O'Dea}, C.~P., {Baum}, S.~A., \& {Morris}, G.~B. 1990, A\&AS, 82, 261

\bibitem[{{Pihlstr{\" o}m} {et~al.}(2003){Pihlstr{\" o}m}, {Conway}, \&
  {Vermeulen}}]{pcv03}
{Pihlstr{\" o}m}, Y.~M., {Conway}, J.~E., \& {Vermeulen}, R.~C. 2003, A\&A,
  404, 871

\bibitem[{{Roettgering} {et~al.}(1997){Roettgering}, {van Ojik}, {Miley},
  {Chambers}, {van Breugel}, \& {de Koff}}]{rvm+97}
{Roettgering}, H.~J.~A., {van Ojik}, R., {Miley}, G.~K., {Chambers}, K.~C.,
  {van Breugel}, W.~J.~M., \& {de Koff}, S. 1997, A\&A, 326, 505

\bibitem[{{Skrutskie} {et~al.}(2006){Skrutskie}, {Cutri}, {Stiening},
  {Weinberg}, {Schneider}, {Carpenter}, {Beichman}, \& {Capps}}]{scs+06}
{Skrutskie}, M.~F., {Cutri}, R.~M., {Stiening}, R., {Weinberg}, M.~D.,
  {Schneider}, S., {Carpenter}, J.~M., {Beichman}, C., \& {Capps}, R. .~M.
  2006, AJ, 131, 1163

\bibitem[{{Srianand} {et~al.}(2010){Srianand}, {Gupta}, {Petitjean},
  {Noterdaeme}, \& {Ledoux}}]{sgp+10}
{Srianand}, R., {Gupta}, N., {Petitjean}, P., {Noterdaeme}, P., \& {Ledoux}, C.
  2010, MNRAS, 1888

\bibitem[{{Stanghellini} {et~al.}(1993){Stanghellini}, {O'Dea}, {Baum}, \&
  {Laurikainen}}]{sobl93}
{Stanghellini}, C., {O'Dea}, C.~P., {Baum}, S.~A., \& {Laurikainen}, E. 1993,
  ApJS, 88, 1

\bibitem[{{Stickel} \& {K\"{u}hr}(1994)}]{sk94}
{Stickel}, M. \& {K\"{u}hr}, H. 1994, A\&AS, 105, 67

\bibitem[{{Stickel} {et~al.}(1996){Stickel}, {Rieke}, {K\"{u}hr}, \&
  {Rieke}}]{srkr96}
{Stickel}, M., {Rieke}, G.~H., {K\"{u}hr}, H., \& {Rieke}, M.~J. 1996, ApJ,
  468, 556

\bibitem[{{Tadhunter} {et~al.}(1993){Tadhunter}, {Morganti}, {di
  Serego-Alighieri}, {Fosbury}, \& {Danziger}}]{tmd+93}
{Tadhunter}, C.~N., {Morganti}, R., {di Serego-Alighieri}, S., {Fosbury},
  R.~A.~E., \& {Danziger}, I.~J. 1993, MNRAS, 263, 999

\bibitem[{Tzanavaris {et~al.}(2007)Tzanavaris, Murphy, Webb, Flambaum, \&
  Curran}]{tmw+06}
Tzanavaris, P., Murphy, M.~T., Webb, J.~K., Flambaum, V.~V., \& Curran, S.~J.
  2007, MNRAS, 374, 634

\bibitem[{Urry \& Padovani(1995)}]{up95}
Urry, C.~M. \& Padovani, P. 1995, PASP, 107, 803

\bibitem[{Vermeulen {et~al.}(2003)Vermeulen, Pihlstr\"{o}m, Tschager, de~Vries,
  Conway, Barthel, Baum, Braun, Bremer, Miley, O'Dea, Roettgering, Schilizzi,
  Snellen, \& Taylor}]{vpt+03}
Vermeulen, R.~C., Pihlstr\"{o}m, Y.~M., Tschager, W., de~Vries, W.~H., Conway,
  J.~E., Barthel, P.~D., Baum, S.~A., Braun, R., Bremer, M.~N., Miley, G.~K.,
  O'Dea, C.~P., Roettgering, H. J.~A., Schilizzi, R.~T., Snellen, I. A.~G., \&
  Taylor, G.~B. 2003, A\&A, 404, 861

\bibitem[{{Webster} {et~al.}(1995){Webster}, {Francis}, {Peterson},
  {Drinkwater}, \& {Masci}}]{wfp+95}
{Webster}, R.~L., {Francis}, P.~J., {Peterson}, B.~A., {Drinkwater}, M.~J., \&
  {Masci}, F.~J. 1995, Nat, 375, 469

\end{thebibliography}

\label{lastpage}

\end{document}